# Tunable narrow linewidth chip-scale mid-IR laser

Euijae Shim[1], Andres Gil-Molina[1], Ohad Westreich[1], Yamac Dikmelik[2], Kevin Lascola[2],

Alexander L. Gaeta[3], Michal Lipson[1,*]

[1]*Department of Electrical Engineering, Columbia University, New York, NY 10027, USA*

[2]*Thorlabs, Inc., Newton, NJ, United States*

[3]*Department of Applied physics and Applied Mathematics, Columbia University, New York, NY 10027, USA*

*Corresponding Author: ml3745@columbia.edu*

**Abstract:** Portable mid-infrared (mid-IR) spectroscopy and sensing applications require widely tunable, narrow linewidth, chip-scale, single-mode sources without sacrificing significant output power. However, no such lasers have been demonstrated beyond 3 μm due to the challenge of building tunable, high quality-factor (Q) on-chip cavities. We demonstrate a tunable, single-mode mid-IR laser at 3.4 μm using a high-Q silicon microring cavity with integrated heaters and a multi-mode Interband Cascade Laser (ICL). We show that the multiple longitudinal modes of an ICL collapse into a single frequency via self-injection locking with an output power of 0.4 mW and achieve an oxide-clad high confinement waveguide microresonator with a loaded Q of $2.8 \times 10^5$. Using integrated microheaters, our laser exhibits a wide tuning range of 54 nm at 3.4 μm with 3 dB output power variation. We further measure an upper-bound effective linewidth of 9.1 MHz from the locked laser using a scanning Fabry-Perot interferometer. Our design of a single-mode laser based on a tunable high-Q microresonator can be expanded to quantum-cascade lasers at higher wavelengths and lead to the development of compact, portable, high-performance mid-IR sensors for spectroscopic and sensing applications.

Chip-scale, widely tunable, narrow-linewidth lasers in the mid-infrared (mid-IR) are critical for a variety of sensing and spectroscopic applications. The 3-5 μm wavelength range is of particular interest for biomedical and environmental gas sensing since this region is the Earth's transparency window where important greenhouse gases and hydrocarbons have their strong absorption signatures[1]. Tunable narrow-linewidth lasers must be developed to achieve fast and precise spectroscopy of gas-phase molecules that are typically narrower than 1 GHz[1]. The goal is to build compact, field-deployable sensors with high performance chip-scale lasers. Such lasers beyond 3 μm wavelength, however, have not been developed up to this point in time. Previous work includes distributed feedback-Quantum/Interband Cascade lasers (DFB-QCL/ICL)[2–4] that can provide high-power and single-mode output for compact mid-IR applications, but wavelength tuning of these lasers is achieved at the cost of significant output power variation related to direct tuning of gain medium. Mid-IR tunable lasers based on external cavities have been demonstrated [5–7] but the bulky free-space external cavities make their miniaturization challenging. V-shaped coupling of two FP cavities inside an ICL for single-mode lasing and wide tunability at 2.8 µm[8] was recently studied to address these challenges. This approach, however, introduces significant complexity, requiring full customization and fabrication of the laser diode.

Building high-performance, chip-scale mid-IR lasers requires the design and development of high quality factor (Q) on-chip cavities in the mid-IR. While chip-scale lasers with frequency tunability and narrow linewidth have been demonstrated using high-Q on-chip cavities[9–11] in the near-IR (NIR), enabling on-chip optical frequency combs[12,13], there is no comparable demonstrations for the mid-IR range. Specifically, engineering tunable high-Q on-chip cavities beyond 3 µm is difficult due to the lack of appropriate tunable transparent materials. Conventional transparent materials in the mid-IR are either not tunable (e.g. fluoride crystallines and chalcogenide glasses[14]) or require customized complex material fabrication such as high-quality molecular-beam epitaxial process (e.g. germanium-on-silicon[15,16]).

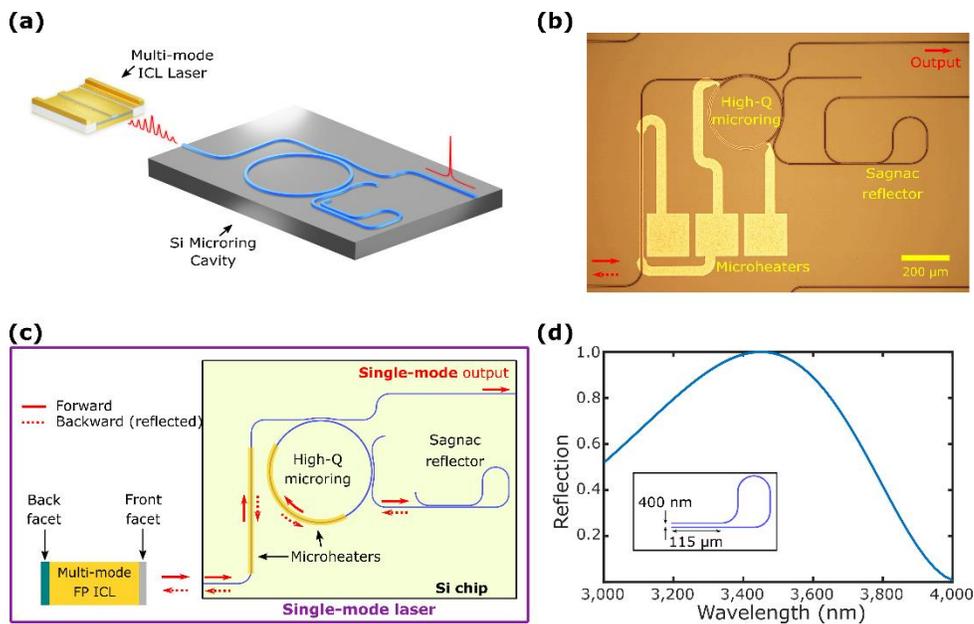

**Figure 1. Single-mode laser platform by integrating a multi-mode ICL laser with a high quality factor (Q) microresonator.** (a) Illustration of a single-frequency mid-IR source, consisting of a multi-mode ICL laser and a silicon microring cavity. (b) Microscope image of the silicon chip with a 150-µm radius high-Q microring (coupling gap of 510 nm), a Sagnac reflector, and aluminum microheaters. (c) Schematic of the single-mode laser. A multi-mode ICL around 3.4 µm with an anti-reflection coating on the front facet. The Sagnac reflector at the drop port provides frequency-selective optical feedback and form an external cavity with the ICL back

facet. Microheaters are placed on top of the bus waveguide and the microring to control the phase of the external cavity and the ring resonance, respectively. (d) Simulated reflection spectrum of the Sagnac loop mirror over wavelengths from 3 μm to 4 μm using Finite-Difference Eigenmode solver. The Sagnac loop mirror reflects about 99% at the drop port at 3.4 μm. A coupler of the Sagnac loop mirror at the drop port consists of a 400 nm gap and a 115 μm length.

Here we demonstrate a tunable single-mode mid-IR laser that uses a high-Q silicon-chip microresonator for self-injection locking of an ICL. Our mid-IR laser design induces spatial collapse of the multi-mode ICL into a single-mode by tuning the ring resonance into one of the ICL modes. To tune the cavity, we employ silicon's thermo-optic effect by leveraging the full CMOS fabrication infrastructure of silicon photonics that provides a compact, chip-scale solution. A schematic of our single-frequency mid-IR laser is shown in Fig1. (a), and the microscope image of the silicon chip is shown in Fig1. (b). A detailed schematic of the operation principle of the single-mode laser is shown in Fig1. (c). In contrast to other laser configurations that leverage the optical feedback to the laser diode or gain chip using backward Rayleigh scattering of the resonators[9–11], we design a microresonator integrated with an external Sagnac reflector where one can control the strength of reflection for optical feedback to the Fabry-Perot (FP) ICL. The reflection of the Sagnac reflector is designed to be above 99 % as shown in Fig1. (d). In this way, we ensure strong optical feedback without relying on the amount of scattering of the resonator[17].

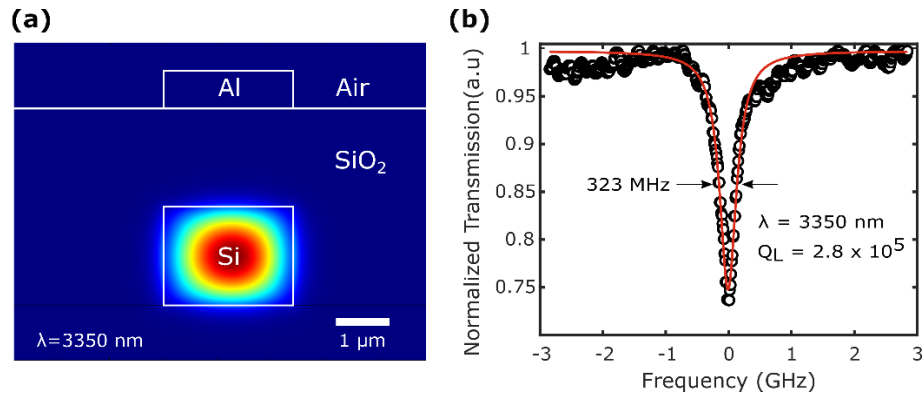

**Figure 2. Mode simulation and normalized transmission spectrum of the fundamental TE mode at 3350 nm.**
(a) Finite Element Mode (FEM) simulation of the fundamental TE mode for a silicon waveguide (2000 nm (H) x 2400 nm (W)) with oxide cladding and aluminum (Al) heater on top for a 150 μm bending radius. (b) Transmission spectrum of the microring at 3350 nm measured using a piezo controlled OPO system. The measured linewidth is 323 MHz, corresponding to the loaded Q of $2.8 \times 10^5$.

We engineer a mid-IR on-chip silicon waveguide ring cavity with high-Q > $10^5$ so that only a single-mode propagates in each direction. In order to simultaneously achieve high Q cavity for high frequency selectivity and tunability, we deposit 2 μm of oxide on the silicon waveguide to isolate the mode from the lossy metal used for the microheaters and design the optical mode to minimally overlap with the absorbing oxide. The silicon waveguide [2000 nm (H) × 2400 nm (W)] is optimized for strong confinement of the fundamental TE mode with a calculated absorption-limited propagation loss of 0.25 dB/cm. We ensure that only a single-mode is fed back to the ICL, despite the fact that the microresonator supports multi-transverse modes, by designing the coupling regions between the microresonator and the Sagnac reflector to be phase-matched only for the fundamental TE. Figure 2 (a) shows the simulated mode profile of the silicon waveguide for a 150-μm radius ring. We experimentally estimate the Q of the microresonator by measuring the transmission spectrum using a tunable optical parametric oscillator (OPO) source near 3.4 μm. Figure 2(b) shows the transmission spectrum of the microresonator with a linewidth of 323 MHz, corresponding to a loaded Q of $2.8 \times 10^5$.

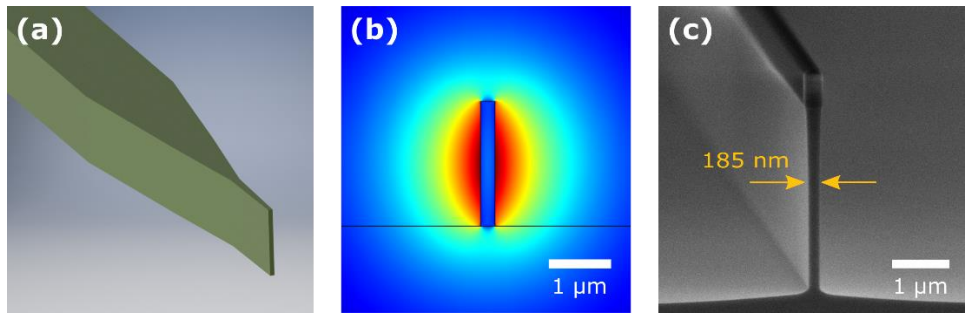

**Figure 3. Design and microscope image of the inverse taper.** (a) Illustration of the inverse taper used to minimize reflections at the output facet. (b) Finite Element Mode (FEM) simulation of the fundamental TE mode of the inverse taper with 2000 nm (H) × 200 nm (W). (c) SEM image of a high aspect-ratio inverse taper structure similar to the one used in our mid-IR laser.

To ensure the dominance of the reflections from the Sagnac reflector (≈ 99%) over undesirable reflections, we use a high aspect-ratio inverse taper. To couple light out of the chip, we design the inverse taper to become delocalized so that the effective index is much lower (≈ 1.459) at 3.4 μm. This lower index decreases Fresnel reflections of a typical silicon-air interface of 30 % down to 3%. Figure 3 (a) illustrates the high-aspect ratio inverse taper, and Fig. 3 (b) shows the mode profile of the inverse taper for fundamental TE mode. We describe chip fabrication in Methods. Figure 3 (c) shows the SEM image of the inverse taper structure fabricated on a silicon chip. Our suspended inverse taper is designed to be robust to any mechanical vibrations by fabricating anchors to the 4 μm of top and bottom oxide cladding. In addition to ensuring low reflections of the output facet, we also design the waveguide to ensure low reflections at the input facet. To decrease Fresnel reflections at the input facet from 30 % to 17 %, we design a Si/SiO$_2$ faceted horn taper with a mode distribution that matches the ICL fundamental mode and clad the taper facet by several microns of $SiO_2$. To prevent leakage of light into the silicon substrate through the 2 μm-thick buried oxide layer, we etch the silicon substrate under the input taper region at the edge of the silicon chip over a length of 5 μm.

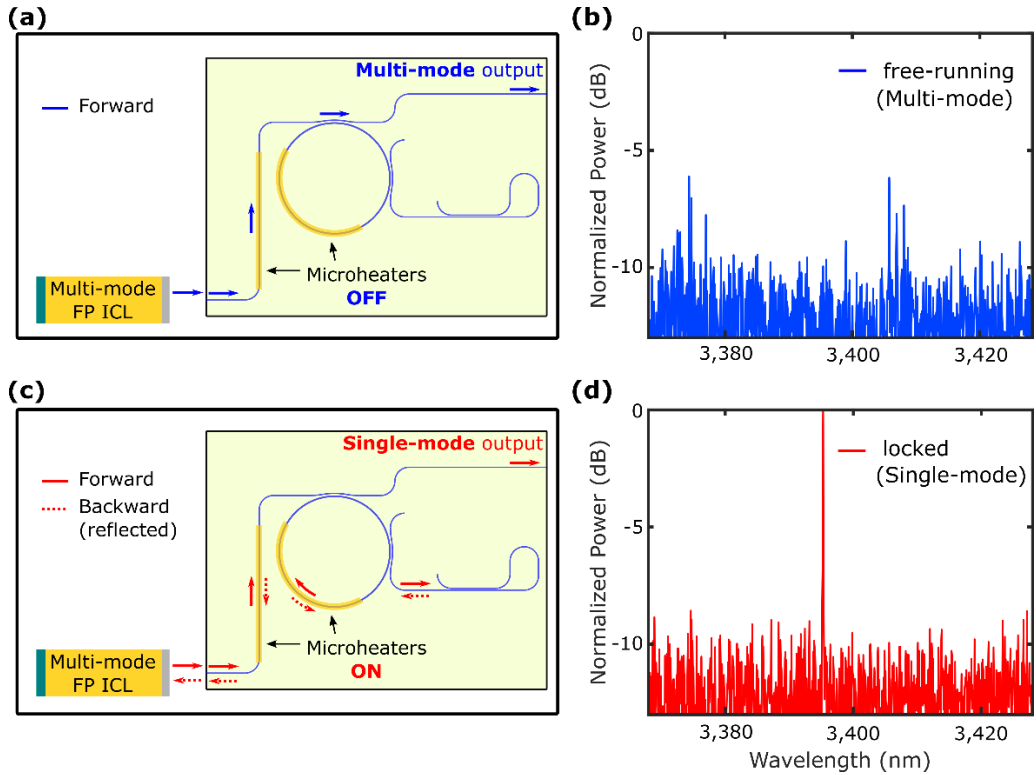

**Figure 4. Self-injection locking of an ICL diode by a microresonator with integrated heaters.** (a) Schematic of the laser setup where microheaters are turned off, and the ring is off-resonance. (b) Optical spectrum of the free-running state laser with multi-mode output. (c) Schematic of the laser setup where microheaters are turned on, and the ring is on-resonance. (d) Optical spectrum of the single-mode locked state laser.

We show single-mode lasing with up to 0.4 mW of output power via self-injection locking of the multi-mode ICL. We observe in Fig 4. (b) the optical spectrum of the unlocked free-running multi-mode ICL when no power is applied to integrated heaters [see Fig 4. (a)], and the ring resonance is off resonance with ICL modes. Figure 4. (d) shows the optical spectrum of injection locked ICL when power is applied to integrated heaters that tune the ring resonance and the phase of the adjacent waveguide [see Fig 4. (c)]. Tuning ensures high reflectivity and constructive interference between the external cavity mode and the ICL mode, leading to the collapse of the ICL multi-mode spectrum into a single-mode. Fig. 4 (d) shows that the locked state power is higher than that of the highest power in the free-running state by at least 6 dB, additional evidence of typical mode collapse[18]. In this high power mode collapse, the power of the ICL is redistributed in favor of the locked mode and therefore can be easily distinguished from a simple frequency filtering effect.

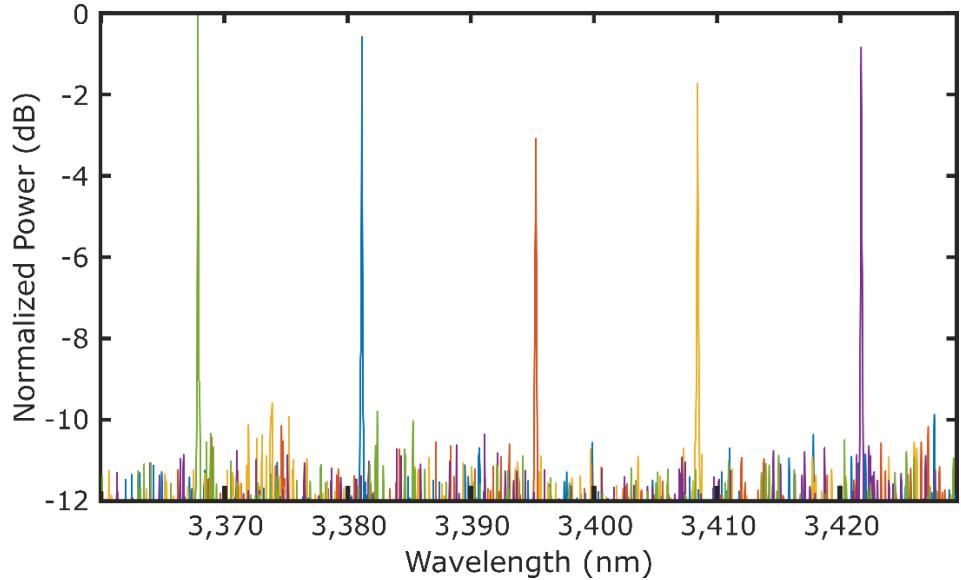

**Figure 5. Tuning of the lasing wavelength in the locked state laser.** Optical spectra showing the wide tunability of the locked state laser of over 54 nm in wavelength near 3.4 μm. Tuning achieved by applying less than 100 mW of electrical power to the integrated heaters.

Our mid-IR laser demonstrates a 54-nm-wide tuning range of lasing wavelength from 3367 nm to 3421 nm. Figure 5 shows the tuning of the single-frequency state in the optical spectrum. We tune the wavelength of a single-frequency state by tuning both the ring (to ensure that the resonance matches with one of the FP ICL modes) and the phase of the adjacent waveguide (to ensure that there is no external cavity net phase accumulated per round trip) by applying less than 100 mW of electrical power to the integrated heaters on top of the bus and ring waveguides. For the spectra of Fig. 5 we apply a diode current of 402 mA (±10%) at the diode temperature of 22°C (±0.5 C). These slight variations of the laser diode operation point allowed us to tune the gain spectrum, enabling wide tunability with low output power variation of 3 dB. Note that our approach for wide tuning of lasing wavelength is controlled by the microresonator resonance frequencies in contrast to the typical temperature tuning in DFB lasers that causes strong changes in the output power.

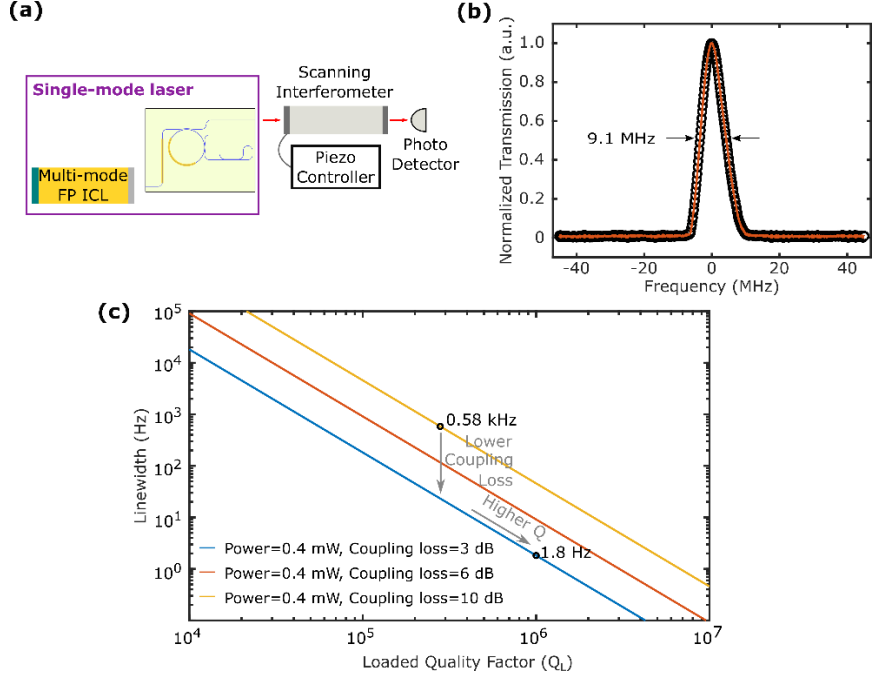

**Figure 6. Measured and calculated linewidth of the locked state laser** (a) Linewidth measurement setup (b) Measured linewidth of 9.1 MHz limited by the resolution of the scanning Fabry-Perot interferometer (c) Calculated fundamental linewidth versus loaded quality factor for 0.4 mW output power. About two orders of magnitude improvement can be achieved with higher $Q_L$ of the microresonator and improved coupling between the ICL and the silicon chip.

We measure an upper-bound for the effective linewidth of 9.1 MHz of the locked laser limited by the resolution of the interferometer. We show the schematic of our experimental setup in Fig. 6 (a) and linewidth measurement in Fig. 6 (b). The locked laser beam is aligned to the interferometer using a pair of irises and collected at a photodetector; the interferometer resonance is scanned with a piezoelectric controller. We estimate the fundamental linewidth (Schawlow-Townes linewidth) of the locked laser to be $\Delta\nu_{locked} \approx 580$ Hz, excluding any source of technical noise and assuming an output power equal to the measured value of 0.4 mW, and coupling loss and quality factor values similar to those in our experiments. Figure 6 (c) shows the calculated fundamental linewidth versus loaded quality factor for different coupling losses between the ICL and the silicon chip. We show that the fundamental linewidth can be further reduced to Hz-level by using a silicon resonator with a $Q$ of $10^6$ and improved coupling loss of 3 dB. Further details of linewidth estimation are discussed in Methods.

In summary, we demonstrate a high-Q tunable microresonator integrated with an off-the-shelf ICL/QCL diode platform for realizing widely tunable, single-mode and narrow-linewidth lasers beyond 3 μm. Our laser design has significant implications for the development of chip-scale spectrometers and gas sensors in the mid-IR since it does not require complex material fabrication or bulky external cavities. In addition, our self-injection locking based approach can be applied to other types of laser cavities as shown in NIR[10,19] such as DFB types at wavelengths up to 8 μm limited by the transparency window of silicon waveguides. Further linewidth reduction of a mid-IR laser diode can be achieved by integrating a higher-Q microresonator like our recently demonstrated Q = $10^6$ air-clad silicon microresonator[20] whose resonance and phase can, in principle, be controlled by placing microheaters on the side of waveguides. Combined with microresonator-based mid-IR combs[21,22] and commercially available higher power ICL/QCLs, our approach paves the way for compact comb-based spectroscopy in the 3-5 μm wavelength region and beyond.

**Methods**

Fabrication

We fabricate the oxide-clad silicon microring resonator and waveguides with integrated heaters using standard CMOS manufacturing processes. We start with a silicon on insulator (SOI) wafer with 2 μm oxide and 2 μm silicon layer thicknesses. To fully etch the 2 μm thick silicon layer with a small coupling gap between the ring and the bus waveguide, we spin ma-N 2410 electron beam (e-beam) resist on the wafer to form a 1.5 μm thick resist layer. To reduce the sidewall roughness, we pattern waveguides using multi-pass e-beam lithography followed by an inductively coupled plasma (ICP) dry etching based on $C_4F_8$ and $SF_6$. After cleaning the wafer with Piranha etchant, we deposit a 2 μm oxide layer using plasma-enhanced chemical vapor deposition (PECVD) for top cladding. We pattern the microheaters by e-beam lithography and sputter layers of 10 nm of titanium and 800 nm of aluminum, followed by a lift-off process in Acetone. We then spin SPR 220-7.0 photoresist and dice the wafer into mm-size chips. After a mild hard bake at 120 ºC, we use xenon fluoride ($XeF_2$) to etch away the silicon substrate near the tapers for efficient coupling.

Experimental setup

We couple the output beam of the ICL to the bus waveguide on a silicon chip using a pair of aspheric lenses that are transparent in the mid-IR. To obtain spectral data, the output of the silicon chip is sent to an optical spectrum analyzer (OSA) after an aspheric lens for beam collimation. The optical spectrum spectral resolution was 4.0 GHz (equivalent to 0.15 nm in wavelength), limited by the OSA. We also co-align an OPO through the chip and send its output to a photodetector connected to an oscilloscope for characterizing the linewidth of the microresonator. For all experiments we measure the output from the mirroring through-port. To measure the linewidth of the locked-state laser, we couple the output beam to a scanning Fabry-Perot interferometer (1.5 GHz FSR, 9 MHz resolution) controlled by a piezo-controller.

Estimation of linewidth

When the laser is self-injection locked to the microresonator, we expect a linewidth reduction from the free-running laser linewidth according to the ref.[10]

$$\Delta\nu_{locked} \approx \Delta\nu_{free} \frac{Q_{ICL}}{Q_{Ring}} \frac{1}{16\eta^2 R(1+\alpha^2)} \qquad (1)$$

where $\Delta\nu_{locked}$ and $\Delta\nu_{free}$ are the linewidth of the laser in the locked and free-running state, respectively. The quality factors $Q_{ICL}$ and $Q_{Ring}$ are those of the ICL and ring resonator, respectively. We estimate the ICL quality factor $Q_{ICL}$ as $Q_{ICL} \approx \frac{2\pi\nu\tau R_{ICL}}{1-R_{ICL}^2}$ where $\tau$ is the ICL cavity round trip and $R_{ICL}$ is the reflection coefficient (amplitude) from the front facet of the ICL. We estimate the $Q_{ICL}$ to be $3.1 \times 10^3$. The parameter R is the reflection from the Sagnac loop mirror that reflects about 10% of light to the laser diode. In addition, we have assumed an ICL linewidth enhancement factor of $\alpha \approx 2.2$, similar to the reference[23]. With these values, the estimated linewidth reduction factor is $\frac{\Delta\nu_{locked}}{\Delta\nu_{free}} \approx 1.3 \times 10^{-3}$. To estimate the linewidth of the locked-state laser, we calculate the linewidth of the free running ICL using the well-known modified Schawlow-Townes formula

$$\Delta\nu_{free} \approx \frac{1+\alpha^2}{4\pi P_{out}} K_{tot} n_{sp}(h\nu) v_g^2 \alpha_m(\alpha_m + \alpha_s) \qquad (2)$$

where $K_{tot}$ is the total enhancement factor, $n_{sp}$ is the population inversion factor, and $v_g$ is the group velocity of the ICL. Considering the output power $P_{out} \approx 0.4\ mW$ and coupling efficiency $\eta \approx 10\ dB$, we estimate the linewidth of the free running state ICL as $\Delta\nu_{free} \approx 0.45\ MHz$, giving the estimated linewidth of the locked state laser $\Delta\nu_{locked} \approx 580\ Hz$.

**Acknowledgements.** This work was supported by Air Force Office of Scientific Research (AFOSR) (FA9550-20-1-0297), Army Research Office (ARO) (grant W911NF-17-1-0016), and Israel's Ministry of Defense – Mission to the U.S. (PO 4441083200). This work was done in part at the City University of New York Advanced Science Research Center NanoFabrication Facility and at the Columbia Nano Initiative (CNI) shared labs at Columbia University in the City of New York. The authors thank Y. Antman, G. Bhatt, I. Datta, and A. Mohanty for helpful discussions.

**Author Contributions.** E.S. designed and fabricated the devices, performed the measurements, and prepared the manuscript. A.G. and O.W. aided the design and measurements. Y.D. and K.L. provided the ICL with an anti-reflective coated facet. M.L. and A.L.G. supervised the research. E.S., A.G., O.W., A.L.G., and M.L. edited the manuscript.

**Competing Interests.** The authors declare no competing financial interests.